\UseRawInputEncoding
\documentclass[preprintnumbers, floatfix, twocolumn,
preprintnumbers, PRD, superscriptaddress,nofootinbib]{revtex4}
\usepackage{eurosym}
\usepackage{amsfonts}
\usepackage{amsmath}
\usepackage{amssymb,epsf}
\usepackage{latexsym}
\usepackage{graphicx,epsfig}
\usepackage{amssymb}
\usepackage{subfigure}
\usepackage{epstopdf}
\usepackage[colorlinks=true,citecolor=blue,linkcolor=blue,urlcolor=black]{hyperref}
\usepackage{color}
\usepackage{float}
\graphicspath{{Images/}}

\renewcommand{\&}{\textup{\symbol{`\&}}}

\begin{document}

\title{Structure formation in mimetic gravity}
\author{Bita Farsi}
\email{Bita.Farsi@shirazu.ac.ir}
\affiliation{Department of Physics, College of Sciences, Shiraz University, Shiraz 71454,
Iran}
\author{Ahmad Sheykhi}
\email{asheykhi@shirazu.ac.ir}
\affiliation{Department of Physics, College of Sciences, Shiraz University, Shiraz 71454,
Iran}
\affiliation{Biruni Observatory, College of Sciences, Shiraz University, Shiraz 71454,
Iran}

\begin{abstract}
We disclose the effects of an extra longitudinal degree of freedom
on the evolution of perturbations in the framework of mimetic
gravity. We consider a flat Friedmann-Robertson-Walker (FRW)
background and explore the linear perturbations by adopting the
spherically symmetric collapse formalism. By suitably choosing the
potential of the mimetic field, we are able to solve the perturbed
field equations in the linear regime and derive the matter density
contrast $\delta_m$ in terms of the redshift parameter $z$. We
observe that $\delta_m$ starts growing at the early stages and as
the universe expands, it grows faster compared to the standard
cosmology. This may due to the extra degree of freedom of the
gravitational field which affects the growth of perturbations.
We observe that in the presence of a mimetic potential,
the growth rate function is smaller than $\Lambda$CDM model in
small redshifts. We then consider the effects of this potential
on the density abundance, the deceleration parameter and jerk
parameter. We find out that mimetic potential can play the role of
dark energy (DE) and affects the dynamics of matter perturbations
and cosmological parameters. We also investigate the mass
function and the number count for the collapsed objects in the
mimetic scenario. We find that the mass function of models with
potential is smaller than model without potential. With decreasing
the role of DE, the mass function start to grow in smaller
redshifts i.e., halo abundance is formed later. It is found that
the more massive structures are less abundant and form at later
times, as it should be in the hierarchical model of structure
formation.

\end{abstract}

\maketitle


\section{Introduction \label{Intro}}
Einstein's theory of General Relativity (GR) is one of the
foundation of modern physics along with the quantum field theory
\cite{Carroll}. Till now, GR has successfully passed various
excremental and observational tests such as bending of light,
gravitational time dilation, gravitational lensing, precession of
the Mercury orbit, etc \cite{Clifford}. Another successful
prediction of GR is the existence of the gravitational waves which
was detected directly from merging of black holes in Laser
Interferometer Gravitational-Wave Observatory (LIGO)
\cite{Abbott}. This detection, a century after the theory was
established by Einstein in $1915$, is a great achievement in
science and opens a new window, gravitational-wave observatory, to
look at the universe. Despite the huge success of GR, it still
suffers to explain some cosmological observations such as the flat
galaxies rotation curves (DM puzzle), the initial and the black
holes singularities and the accelerated expansion of the universe
(DE puzzle). Therefore, either alternative theories of gravity or
new components of matter/energy have been proposed to explain the
observed phenomenons.

Modified gravity theories have achieved great development and
performance in explanation some unsolved problems of GR. One of
the modification to GR is the mimetic theory of gravity which was
suggested as a new explanation for the DM puzzle \cite{Mim1}. In
this scenario, a mimetic scalar field $\Phi$ is introduced which
is not dynamical by itself, nevertheless it makes the longitudinal
degree of freedom of the gravitational field dynamical. This
dynamical longitudinal degree of freedom of the gravitational
field can play the role of mimetic pressureless DM \cite{Mim1}. A
modified version of mimetic gravity can address the cosmological
singularities \cite{MimCos1} as well as the singularity in the
center of a black hole \cite{Cham3}. Besides, it has been
confirmed that the original setting of the mimetic theory predicts
that gravitational wave (GW) propagates at the speed of light,
ensuring agreement with the results of the event GW170817 and its
optical counterpart \cite{Sunny1,Sunny2,Sherafati}. It has also
been shown that this theory can explain the flat rotation curves
of spiral galaxies without needing to particles DM
\cite{MimMOND,SheyGr}. Mimetic theory of gravity has arisen a lot
of enthusiasm in the past few years both from the cosmological
viewpoint \cite{MimCos,
Baf,Dutta1,Shey3,Sep,Mat,Leb,MimCos2,Gorj1,Gorj2,Gorj3,Gorj4,Fir,Cham2,Russ,Ces,Vic,Fir2,Arroja,Mansoori}
as well as black holes physics \cite{Der,Myr1,Myr2,Alex,
Chen,Nash3,jibril,Bra,Yunlong1,Yunlong2,Nash1,Cham4,Gorji2,Shey1,Shey2,NashNoj,ACham,YZ,Bakh,Nash4}.
The investigations on the mimetic theory of gravity were also
generalized to $f(R)$ gravity
\cite{Odin0,Odin1,Oik1,Oik2,Oik3,MyrzSeb,OdinP,OdinOik,Odin3,Odinfr1,Odinfr2,Bhat,Adam,Jing}
as well as Gauss-Bonnet gravity \cite{OdinGB,Oik,Zh1,Zh2,Paul}. In
particular, a unified description of early inflation and late-time
acceleration in the context of mimetic $f(R)$ gravity was
established in \cite{Odin2}. It has been confirmed that in the
background of mimetic $f(R)$ gravity, the inflationary era can be
realized \cite{Odin2}.

Observations confirm that almost 26\% of the energy budget of the
universe correspond to the DM sector, while around 69\% constitute
the DE \cite{Yang}. There are enough evidences that support the
existence of DM and DE in the universe \cite{Seljak}. DE has
became relevant only more recently and is presumed to be a smooth
component with a negative pressure. DE has anti-gravity feature
and pushes the universe to accelerate and therefore is responsible
for the acceleration of the cosmic expansion (for a recent review
on DE and DM see \cite{Oks}). DM plays two important roles in the
universe evolution. First it provides an enough gravity for the
rotation of the spiral galaxies as well as cluster of galaxies.
Second, it plays a crucial role in the growth of perturbation and
structure formation in the early stage of the universe. The latter
is due to the fact that DM has no electromagnetism interaction (no
photon radiation) while it enjoys the gravitational interaction.
As a result, DM begins to collapse into a complex network of DM
halos, while ordinary matter continues its collapse due to the
photon radiation and finally settles down into the well potential
of the DM. Without DM, the epoch of galaxy formation would occur
substantially later in the universe than is observed.

Understanding the origin and physics of the growth of
perturbations in galaxies and clusters of galaxies, has been one
of the main challenges of the modern cosmology. This is of great
importance, since these perturbations eventually lead the large
scale structure of the universe. It is a general belief that the
large scalae structure of the universe such as galaxies and
clusters of galaxy arise from gravitational instability that
amplifies very small initial density fluctuations during the
universe evolution. Such fluctuations then grow slowly over time
until they get robust enough to be detached from the background
expansion. Finally, they collapse into gravitationally-bound
systems such as galaxies and clusters of galaxy. In other words,
the primordial collapsed regions serve as the initial cosmic seeds
for which the large scale structures are developed
\cite{Peebles,White}.

An appropriate approach to investigate the growth of perturbations
and structure formation is the so called Top-Hat Spherical
Collapse (SC) model. The results of the observational data
\cite{Huterer} show that most of the structures are formed from
nonlinear evolution of perturbations in the dark age period
($10<z<100$) \cite{Miralda}. The dark age is the period between
the time when the cosmic microwave background was emitted and the
time when the evolution of structure in the universe led to the
gravitational collapse of objects, in which the first galaxies
were formed. The simplest analytical approach to study non-linear
structure formation is the SC model \cite{Abramo}. In this model,
one considers a uniform and symmetrical spherical perturbation in
an expanding background. The symmetry of this model allows us to
treat a spherical perturbation in a FRW universe. In other words,
we can describe the growth of perturbations in a spherical region
using the same Friedmann equations for the underlying theory of
gravity \cite{Planelles}. According to SC model, at early times,
primordial spherical overdense regions expand along the Hubble
flow. Since the relative overdensity of the overdense region with
respect to the background is small, the linear theory is enough to
study their evolution. At a certain point, gravity starts
dominating and overcome the expansion rate. Eventually, the sphere
reaches a maximum radius and completely detaches from the
background expansion. The following phase is represented by the
collapse of the sphere under its own self-gravity. The exact
process of collapse due to gravitational instability depends
strongly on the dynamics of the background Hubble flow
\cite{Naderi}. The effects of DE on the structure formation has
been investigated in various scenarios \cite{SDutta2}.  In the
context of mimetic gravity, the cosmological perturbations has
been investigated in \cite{Matsumoto}. Our work differs from
\cite{Matsumoto} in that instead of using the Newtonian gauge, we
use SC formalism to examine the evolution of perturbations.

The outline of this paper is as follows. In section \ref{RMC}, we
provide a review on mimetic gravity and derive the corresponding
Friedmann equations in the context of mimetic cosmology. In
section \ref{GP}, using the spherically collapse approach, we
explore the growth of matter perturbation in the background of
flat mimetic cosmology. In section \ref{MN}, we study the
mass function and number count of the collapsed objects in the
mimetic scenario. The last section is devoted to conclusion and
discussion.
 \section{Mimetic cosmology\label{RMC}}
According to the mimetic theory of gravity, the physical metric
$g_{\mu \nu}$  is related to the auxiliary metric $\tilde{g}_{\mu
\nu}$ and scalar field $\Phi$ through a conformal transformation
\cite{Mim1}
\begin{equation}
g_{\mu \nu}=\tilde{g}_{\mu
\nu}\tilde{g}^{\alpha\beta}\partial_{\alpha}\Phi\partial_{\beta}\Phi,
\label{tran}
\end{equation}
where the physical metric $g_{\mu \nu}$ is invariant under the
conformal transformations of the auxiliary metric $\tilde{g}_{\mu
\nu}$, namely, under $\tilde{g}_{\mu \nu}\rightarrow \Omega^2
\tilde{g}_{\mu \nu}$, we have  $g_{\mu \nu}\rightarrow g_{\mu
\nu}$.

The action of mimetic gravity in the presence of a mimetic
potential is given by \cite{Dutta1}
\begin{equation}\label{action}
S=\int d^{4}x
\sqrt{-g}\left[\dfrac{R}{2\kappa^{2}}+\dfrac{\lambda}{2}
\left(\partial^{\mu}\Phi\partial_{\mu}\Phi-1\right)-V(\Phi)+\mathcal{L}_{\rm
m}\right],
\end{equation}
where $R$ is the Ricci scalar, $\mathcal{L}_{\rm m}$ is the
Lagrangian of matter, $g$ is the determinant of the physical
metric, and $\lambda$ is a Lagrange multiplier. Here $V(\Phi)$ is
an arbitrary function of the scalar field $\Phi$, and the factor
$1/2$ in front of the Lagrange multiplier $\lambda$ is introduced
for latter convenience. Such a model has been shown to provide an
economical way of reproducing a number of simple and
well-motivated cosmological scenarios, relevant for both early-
and late-time cosmology, without the need for neither an explicit
DM nor dark fluid \cite{MimCos}. Throughout this work, we take
$\kappa^{2}=8\pi G=1$, for simplicity. By vary the above action
with respect to the physical metric $g_{\mu\nu}$, one can derive
the field equations as \cite{Dutta1}
\begin{equation}
G_{\mu\nu}+\lambda\partial_{\mu}\Phi\partial_{\nu}\Phi+g_{\mu\nu}V(\Phi)=T_{\mu\nu},
\label{field eq1}
\end{equation}
where $G_{\mu\nu}$ is the Einstein tensor and $T_{\mu\nu}$ is the
energy momentum tensor of the usual matter. Variation of the
action (\ref{action}) with respect to Lagrange multiplier $\lambda$ yields
\begin{equation}
g^{\mu\nu}\partial_{\mu}\Phi\partial_{\nu}\Phi=1. \label{con}
\end{equation}
This constraint, which is consistent with conformal transformation
(\ref{tran}), implies that the scalar field is not dynamical by
itself, however, it encodes a new longitudinal degree of freedom
to the gravitational field \cite{Mim1}. By varying action
(\ref{action}) with respect to the mimetic field $\Phi$, we find
\cite{Dutta1}
\begin{equation}
\nabla^{\mu}\left(\lambda\partial_{\mu}\Phi\right)+\dfrac{dV}{d\Phi}=0. \label{field
eq2}
\end{equation}
This equation can also be derived by taking the covariant
derivative of Eq. (\ref{field eq1}) and using the fact that
$\nabla^{\mu} G_{\mu\nu}=0$, together with the continuity equation
$\nabla^{\mu} T_{\mu\nu}=0$. We also assume the background is
homogeneous, isotropic and spatially flat with line elements
\begin{equation}
ds^{2}=dt^{2}-a^{2}(t)(dx^2+dy^2+dz^2), \label{metric}
\end{equation}
where $a(t)$ is the scale factor of the universe and we have taken
the spacetime signature as $(+,-,-,-)$. For homogenous cosmology,
the mimetic scalar field is only a function of time, i.e.,
$\Phi=\Phi(t)$. Inserting into condition (\ref{con}) immediately yields
\begin{equation}
\dot\Phi^{2}=1.
\end{equation}
Integrating (assuming $\dot\Phi>0$), we get $\Phi=t$, where we
have also set the integration constant equal to zero. We further
assume the energy content of the universe is in the form of
perfect fluid with energy-momentum tensor $T_{\nu}^{\mu}={\rm
diag}\left(\rho,-p,-p,-p\right)$, where $\rho$ and $p$ are,
respectively, the energy density and pressure of the fluid.
Conservation of energy-momentum tensor, $\nabla^{\mu}
T_{\mu\nu}=0$, in the background of FRW universe leads to
\begin{equation}
\dot{\rho}+3H(\rho+p)=0. \label{con2}
\end{equation}
So the energy density of the pressureless matter can be written as
$\rho_{m}=\rho_{m,0}a^{-3}$. The cosmological field equations can
be derived by substituting the metric (\ref{metric}) and the
energy momentum tensor in the field Eqs. (\ref{field eq1}) and
(\ref{field eq2}). We find \cite{Dutta1}
\begin{eqnarray}
&&3H^{2}=\rho-\lambda-V,\label{Fried1}\\
&&3H^{2}+2\dot{H}=-(p+V),\label{Fried2}\\
&&\dot{\lambda}+3H\lambda+\dfrac{dV}{d\Phi}=0,\label{Fried3}
\end{eqnarray}
where $H\equiv \dot a/a$ is the Hubble parameter. Let us note that
we can distinguish three sectors for the energy components. First
is the usual matter with energy density $\rho$ and pressure $p$.
The second is the mimetic field which plays the role of DM and
contributes trough $\lambda$ in the field equations. And third the
mimetic potential, $V$, which plays the role of DE.  The first
Friedmann equation (\ref{Fried1}), using the definition of density
parameters, can be rewritten
\begin{eqnarray}
&&\Omega_{m}+\Omega_{\lambda}+\Omega_{V}=1,
\label{Omega}
\end{eqnarray}
\begin{eqnarray}
&&\Omega_{m}
\equiv\frac{\rho_{m}}{3H^2},\label{Omega1}\\
&&\Omega_{dm}\equiv\Omega_{\lambda}\equiv\frac{-\lambda}{3H^2}\label{Omega2},\\
&&\Omega_{de}\equiv\Omega_{V}\equiv\dfrac{-V}{3H^2},\label{Omega3}
\end{eqnarray}
Using the fact that $\Phi=t$, Eq. (\ref{Fried3}), can be written as
\begin{equation}
\frac{d \lambda}{d\ln a}+3\lambda+\frac{d V}{d\ln a}=0,
\label{Va0}
\end{equation}
Thus, if we make the following ansatz for the potential,
\begin{equation}
V(a)=-\frac{\beta}{1+\beta}\lambda (a),
\end{equation}
then we can solve analytically Eq. (\ref{Va0}) to arrive at
\begin{equation}
\lambda(a)=\lambda_{0} a^{-3(1+\beta)}, \label{lambda}
\end{equation}
where $\beta$ is a dimensionless constant and $\lambda_{0}$ is the
present value of $\lambda$, and we have chosen $a_0=a(t=t_0)=1$.
Hence the explicit form of the mimetic potential is given by
\begin{equation}
V(a)=-\dfrac{\beta \lambda_{0}}{\beta+1} a^{-3(1+\beta)},
\label{Va}
\end{equation}
In the absence of the mimetic potential ($\beta=0=V$), from Eq.
(\ref{lambda}) one can see that $\lambda=\lambda_{0}a^{-3}$,
namely the field mimics DM \cite{Mim1}.

As we shall see the mimetic potential indeed plays the role of DE
and can lead to the acceleration of the cosmic expansion. Besides,
in our model, we assumed the mimetic potential (which plays the
role of DE) is a function of $\lambda$ (which plays the role of
DM). This implies an interaction between DE and DM.
The Hubble expansion rate can be obtained via Eq. (\ref{Fried1}) as
\begin{equation}
H^2=\dfrac{\rho_{m,0}(1+\beta)-\lambda_{0}a^{-3\beta}}{3(1+\beta)a^{3}}. \label{Hubble}
\end{equation}
So we can define the normalized Hubble parameter as
\begin{eqnarray} \label{Ez}
&&E(z)=\dfrac{H(z)}{H_{0}}=\sqrt{ \dfrac{\rho_{m,0}(1+\beta)-\lambda_{0}(1+z)^{3\beta}}{3H_{0}^{2}(1+\beta)(1+z)^{-3}}},\nonumber\\
&&=\sqrt{\dfrac{\left[ (1+\beta)+\dfrac{\Omega_{\lambda,0}}{\Omega_{m,0}}(1+z)^{3\beta}\right](1+z)^3}{(1+\beta)+\dfrac{\Omega_{\lambda,0}}{\Omega_{m,0}}}}.
\end{eqnarray}
The evolution of the normalized Hubble parameter versus $z$ for
different values of $\beta$ is plotted in Fig. 1.
\begin{figure}[t]
\epsfxsize=7.5cm \centerline{\epsffile{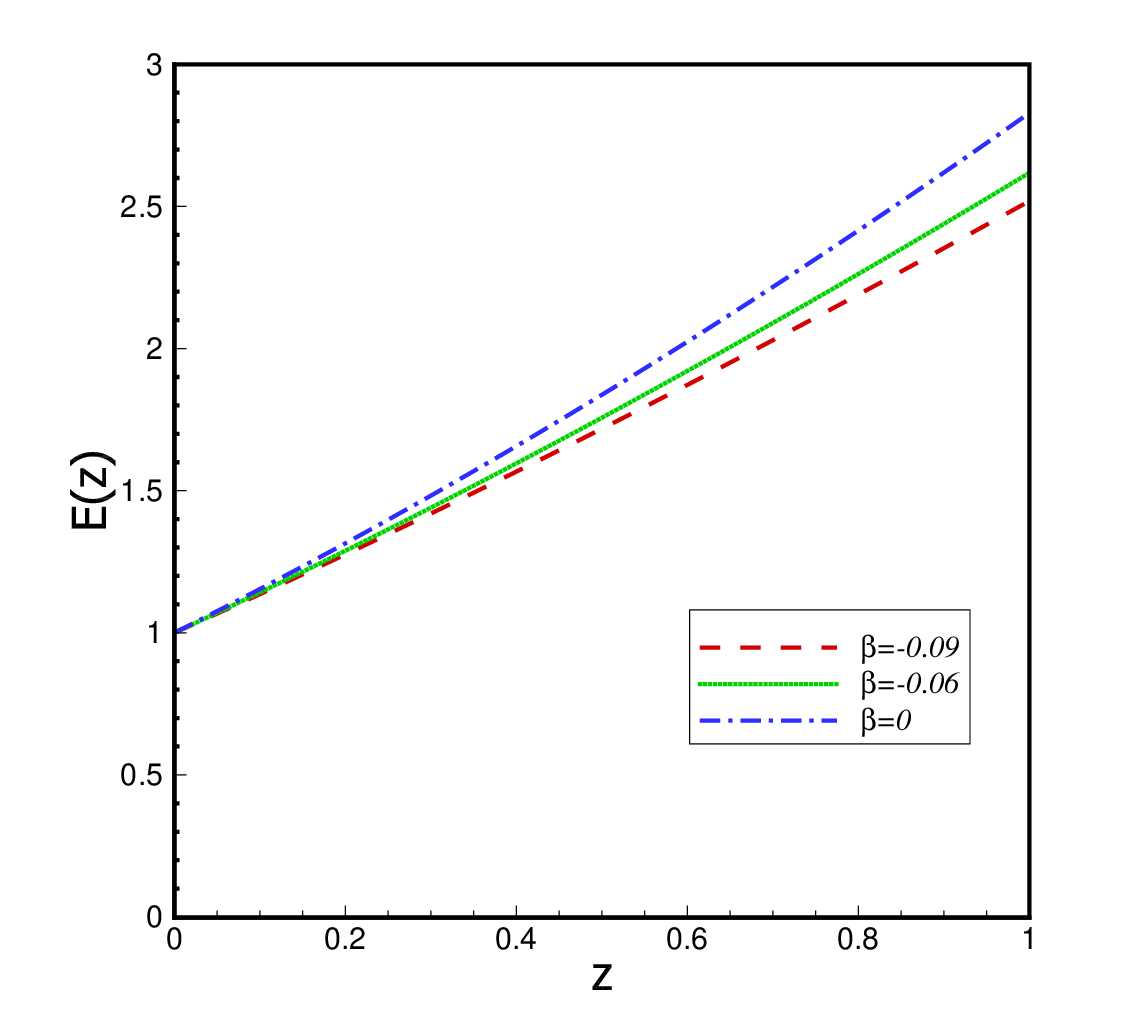}} \caption{The
behavior of the normalized Hubble rate for different values of
$\beta$.} \label{fig1}
\end{figure}

In Fig. 2, we have plotted the evolution of the density abundance
$\Omega_{m}$, defined as
\begin{eqnarray}
\Omega_{m}(z)&=&\dfrac{\rho_{m}}{3H^2}=\dfrac{(1+\beta)\rho_{m,0}}{(1+\beta)\rho_{m,0}-\lambda_{0}(1+z)^{3\beta}},\nonumber\\
&=&\dfrac{(1+\beta)\Omega^{0}_{m}}{(1+\beta)\Omega^{0}_{m}+\Omega^{0}_{\lambda}(1+z)^{3\beta}}.
\label{Omegam}
\end{eqnarray}
\begin{figure}[t]
\epsfxsize=7.5cm \centerline{\epsffile{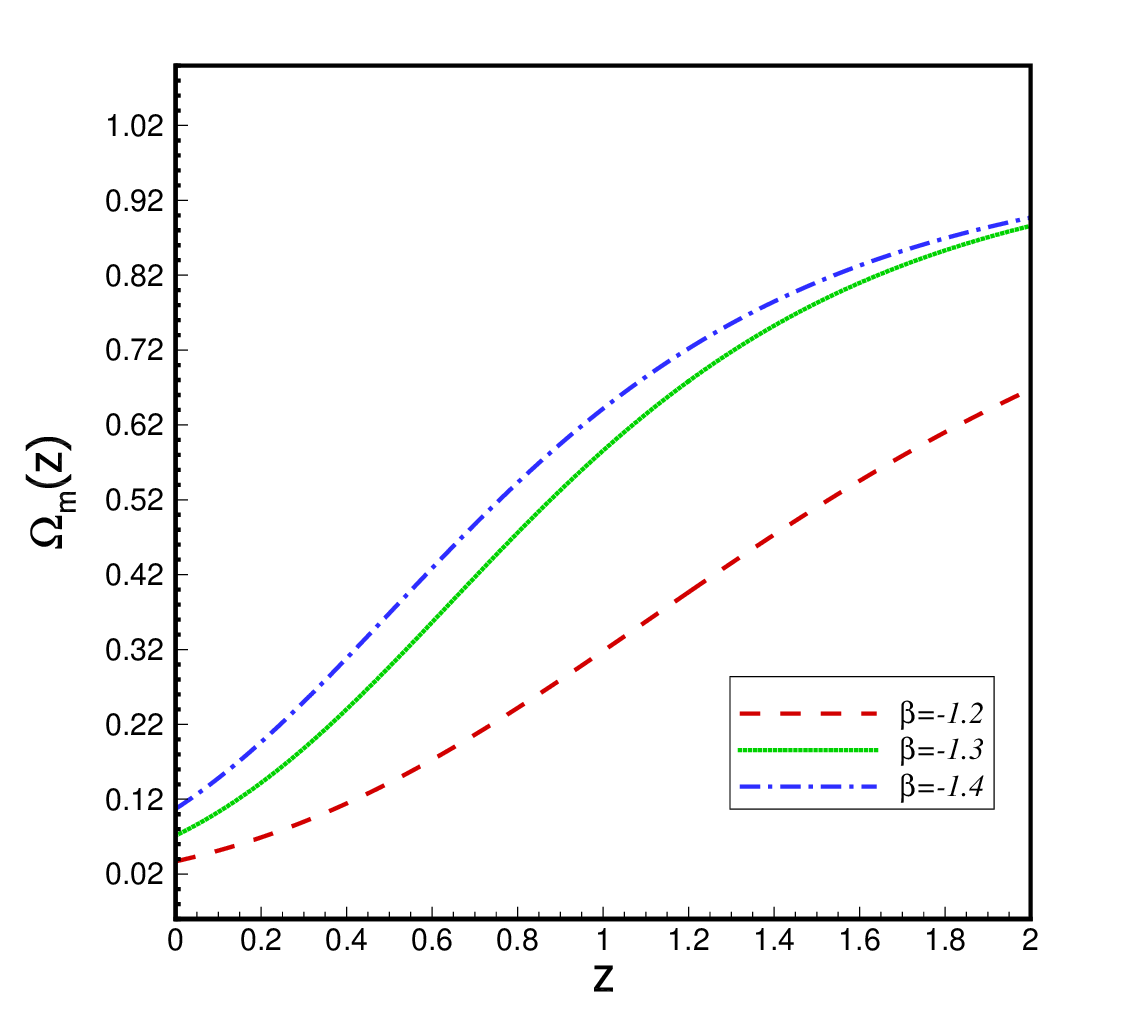}} \caption{The
evolution of the matter density abundance as a function of
redshift $z$ for different values of $\beta$. } \label{fig2}
\end{figure}
As we can see from Fig. 2, the matter density abundance with
different $\beta$ parameters have the same behavior at first, i.e.
all graphs are reduced by decreasing $z$. In addition for smaller
values of $\beta$ parameter,
the density abundance drops faster.\\
We can also write the explicit expression for $\Omega_{\lambda}$,
as
\begin{eqnarray}
\Omega_{\lambda}(z)&=&=\dfrac{-\lambda}{3H^2}=\dfrac{-\lambda_{0}(1+\beta)(1+z)^{3\beta}}{(1+\beta)\rho_{m,0}-\lambda_{0}(1+z)^{3\beta}},\nonumber\\
&=&\dfrac{(1+\beta)\Omega^{0}_{\lambda}(1+z)^{3\beta}}{(1+\beta)\Omega^{0}_{m}+\Omega^{0}_{\lambda}(1+z)^{3\beta}}.
\label{Omegalambda}
\end{eqnarray}
\begin{figure}[t]
\epsfxsize=7.5cm \centerline{\epsffile{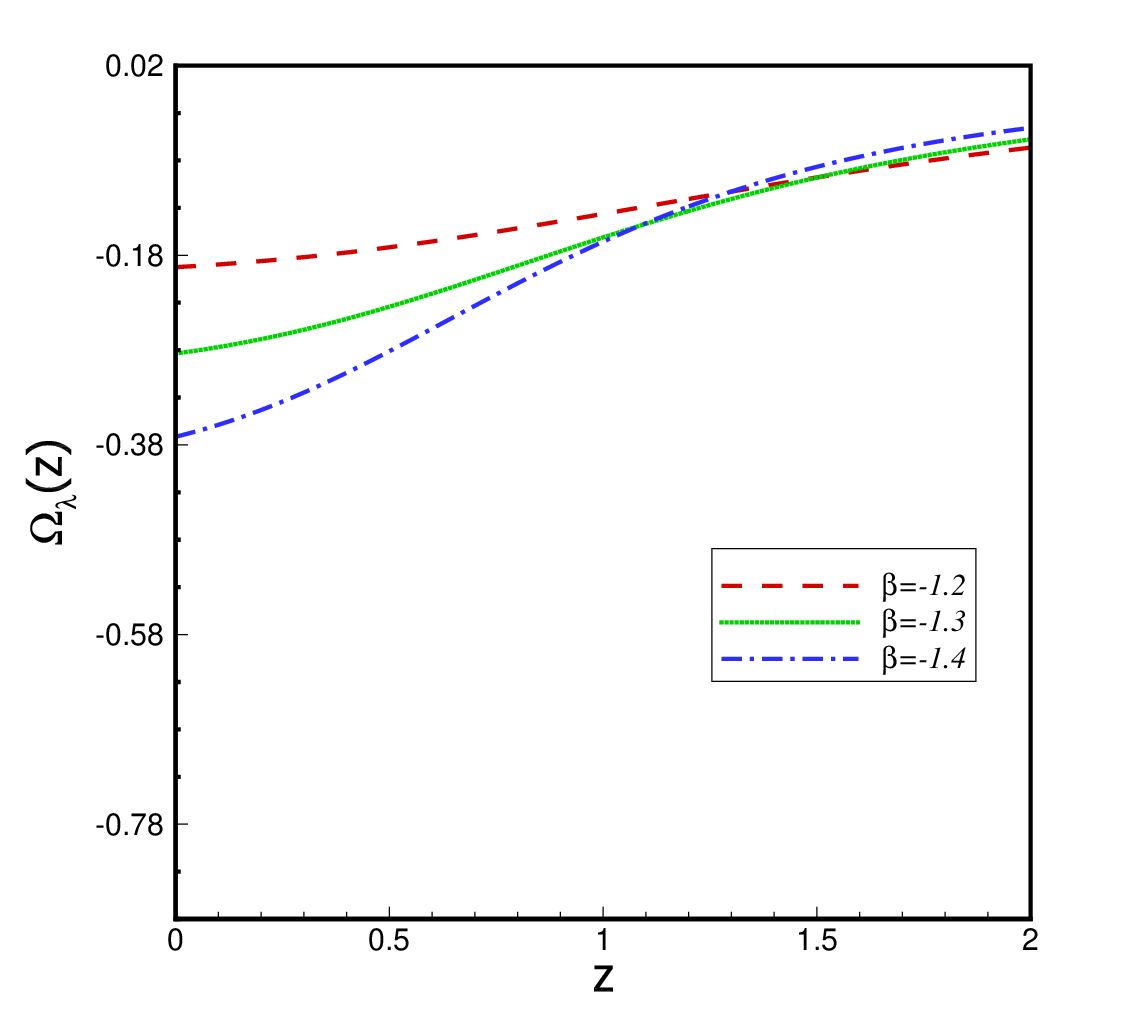}} \caption{The
evolution of the DM density abundance in terms of the redshift $z$
for different values of $\beta$.} \label{fig3}
\end{figure}
In a similar way, the evolution of the density abundance
$\Omega_{V}$, is given by
\begin{eqnarray}
\Omega_{V}(z)&=&\dfrac{-V}{3H^2}=\dfrac{\beta\lambda_{0}(1+z)^{3\beta}}{(1+\beta)\rho_{m,0}-\lambda_{0}(1+z)^{3\beta}},\nonumber\\
&=&-\dfrac{\beta\Omega^{0}_{\lambda}(1+z)^{3\beta}}{(1+\beta)\Omega^{0}_{m}+\Omega^{0}_{\lambda}(1+z)^{3\beta}}.
\label{OmegaV}
\end{eqnarray}
\begin{figure}[t]
\epsfxsize=7.5cm \centerline{\epsffile{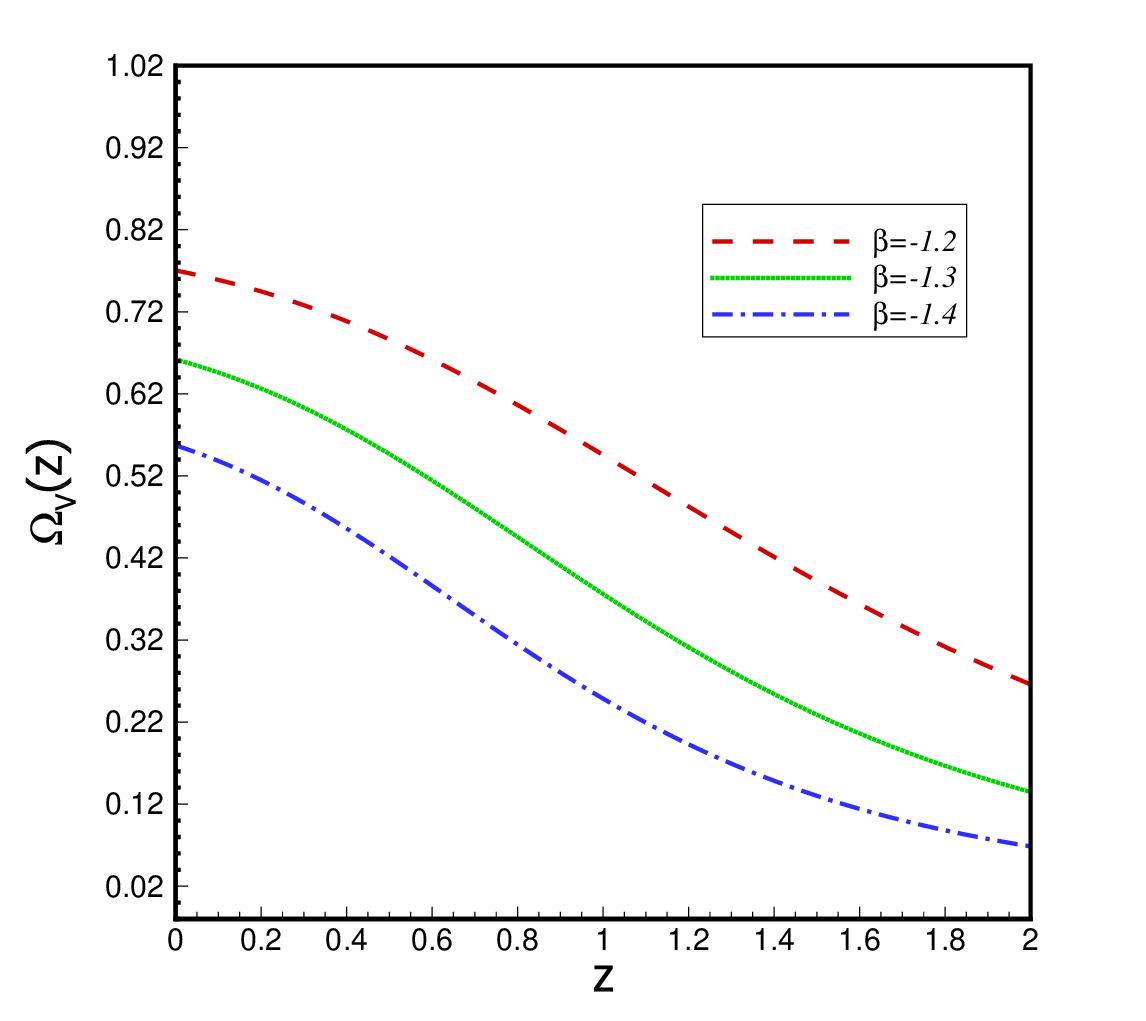}} \caption{The
evolution of the DE density abundance in terms of the redshift $z$
for different values of $\beta$.} \label{fig4}
\end{figure}
As we can see from Figs 2, 3 and 4, in the framework of mimetic
gravity, the different potential suggest different values for
$\Omega^{0}_{m}$, $\Omega^{0}_{\lambda}$ and $\Omega^{0}_{V}$.
Comparing to standard values of the density parameters, i.e.
$\Omega^{0}_{m}=0.05$, $\Omega^{0}_{\lambda}=0.26$ and
$\Omega^{0}_{V}=0.69$ , we find that the best value for the
$\beta$ parameter is $\beta\simeq-1.3$, which corresponds
to $\Lambda C D M$ model.

The deceleration parameter in terms of the redshift can be written
as
\begin{equation}
q=-1-\dfrac{\dot{H}}{H^2}=-1+\dfrac{(1+z)}{H(z)}\dfrac{dH(z)}{dz},
\label{deceleration}
\end{equation}
Using Eq. (\ref{Hubble}) the deceleration parameter is given by
\begin{equation}
q(z)=\dfrac{1}{2}\left[\dfrac{(\beta+1)\rho_{m,0}-(3\beta+1)
\lambda_{0}(1+z)^{3\beta}}{(\beta+1)\rho_{m,0}-\lambda_{0}(1+z)^{3\beta}}
\right]. \label{deceleration2}
\end{equation}
Let us note that for either $\beta=0$  or $\lambda_0=0$, we have
$q=1/2$, which is the result of standard cosmology in the presence
of dust. We have plotted the behavior of the deceleration
parameter $q(z)$ for different $\beta$ parameter in Fig. 5. We
observe that the universe experiences a transition from
decelerating phase ($z>z_{tr}$) to accelerating phase
($z<z_{tr}$), where $z_{tr}$ is the redshift at the transition
point. Also we can see that with decreasing $\beta$, the phase
transition between deceleration and acceleration take place at
lower redshifts. Note that $q(z)$ has a phase transition, for
different values of $\beta$, form deceleration ($q>0$) to
acceleration ($q<0$) phase, around $z\approx 0.6$ which is
compatible with observational data.\\
\begin{figure}[t]
\epsfxsize=7.5cm \centerline{\epsffile{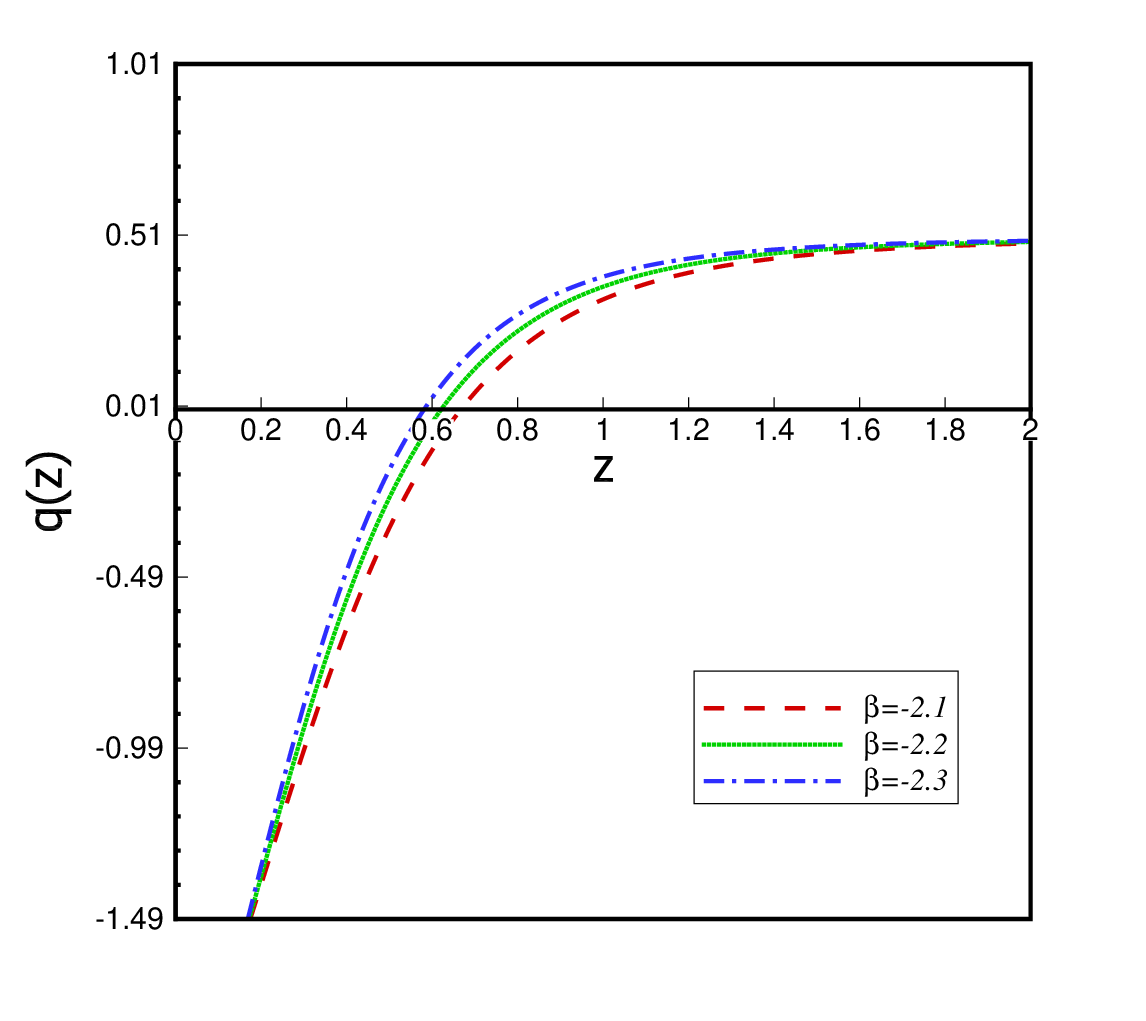}} \caption{The
behavior of the deceleration parameter as a function of redshift
for different $\beta$.}. \label{fig5}
\end{figure}
Another quantity which is helpful in understanding the
phase transitions of the universe expansion is called the jerk
parameter. This is a dimensionless quantity obtained by taking the
third derivative of the scale factor with respect to the cosmic
time, provides a comparison between different DE models and the
$\Lambda$CDM $(j=1)$ model. The jerk parameter is defined as
\cite{Visser, Rapetti, Mamon}
\begin{equation}
j=\dfrac{1}{aH^{3}}\dfrac{d^{3}a}{dt^{3}}=q(2q+1)+(1+z)\dfrac{dq}{dz}.
\label{jerk}
\end{equation}
As we know the Hubble parameter, the deceleration parameter and
the jerk parameter are purely kinematical, since they are
independent of any gravity theory, and all of them are only
related to scale factor $a$ or redshift $z$. For the $\Lambda$CDM
model, the value of $j$ is always unity. A non-$\Lambda$CDM model
occurs if there is any deviation from the value of $j=1$. This is
similar as deviation from the EoS parameter $\omega=-1$, for other
DE models. Using Eq. (\ref{deceleration}) in Eq. (\ref{jerk}), we
can find the jerk parameter in our model. From Fig. 6 we observe
the jerk parameter at early time ($z>1$) tends to a $\Lambda CDM$
model ($j=1$).
\begin{figure}[t]
\epsfxsize=7.5cm \centerline{\epsffile{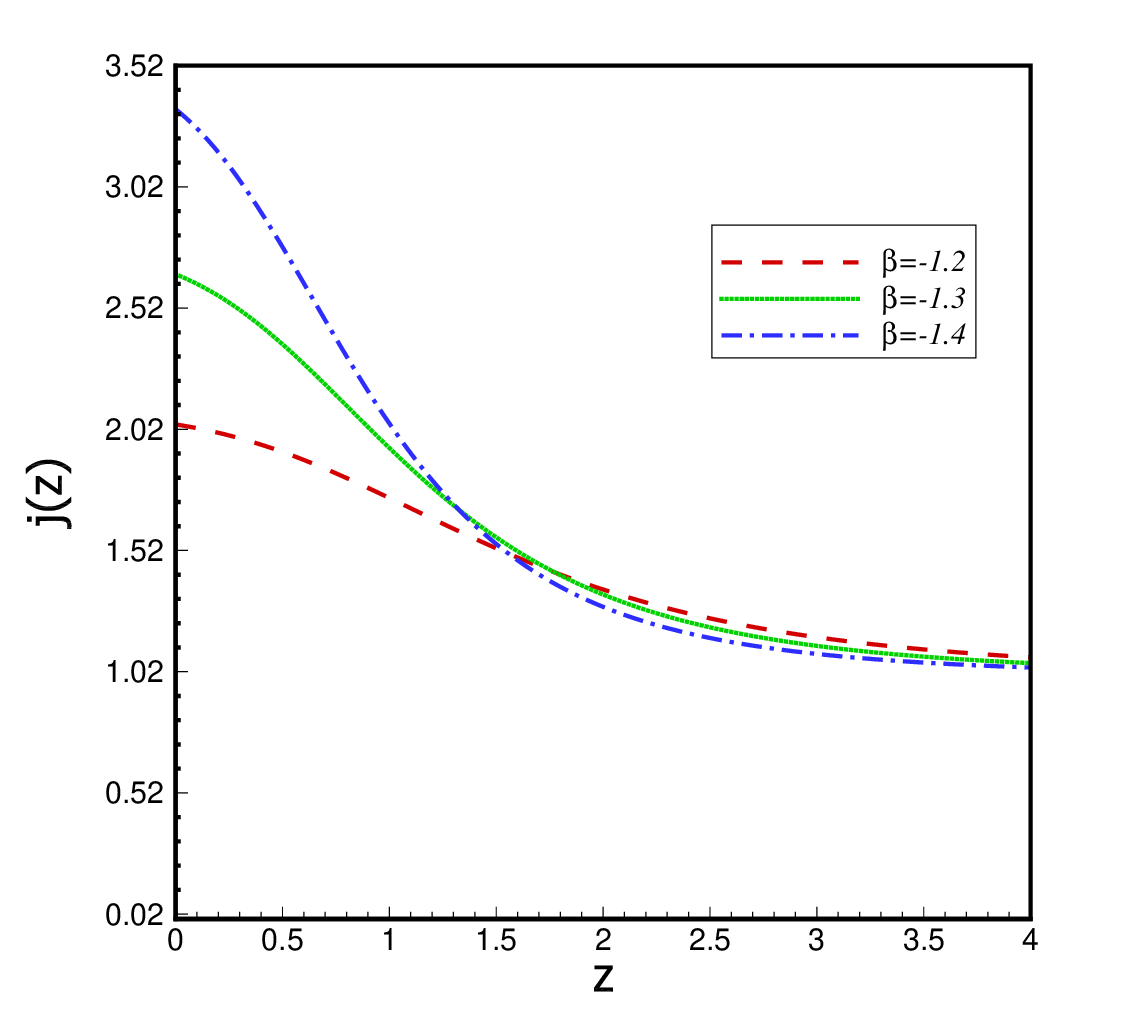}} \caption{The evolution of
jerk parameter with respect to redshift for different values of $\beta$ parameter.} \label{fig6}
\end{figure}\
Finally, we can plot the mimetic potential given in Eq.(\ref{Va})
as a function of redshift parameter. From Fig. 7, we observe that
as the universe expands, the potential increases from the past to
the present.
\begin{figure}[t]
\epsfxsize=7.5cm \centerline{\epsffile{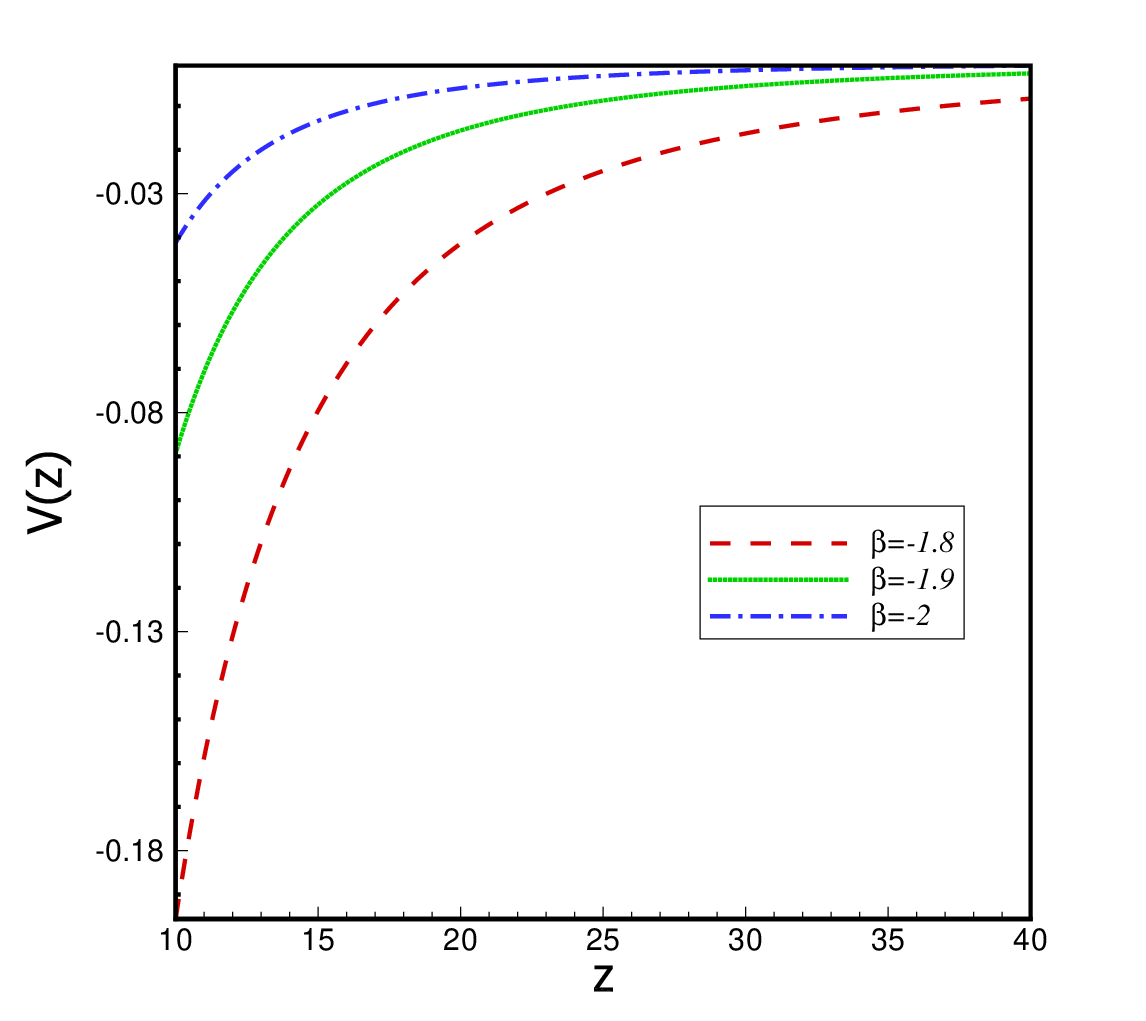}} \caption{The
behavior of the mimetic potential in terms of the redshift
parameter for different $\beta$.} \label{fig7}
\end{figure}\
\section{Growth of perturbation in mimetic cosmology\label{GP}}
We consider a universe filled with pressureless matter,
$p=p_{m}=0$. In this case Eq.(\ref{con2}) reads
\begin{equation}
\dot{\rho}_{m}+3H\rho_{m}=0,
 \label{continuty1}
\end{equation}
which has a solution of the form $\rho_{m}= \rho_{m,0}a^{-3}$,
where $\rho_{m,0}$ is the energy density at the present time. In
order to study the growth of perturbations, we consider a
spherically symmetric perturbed cloud of radius $a_{p}$, and with
a homogeneous energy density $\rho_{m}^{c}$. The SC model
describes a spherical region with a top-hat profile and uniform
density so that at any time $t$, we can write
$\rho_{m}^{c}(t)=\rho_{m}(t)+\delta\rho_{m}$\cite{ziaei}. If
$\delta\rho_{m}>0$ this spherical region will eventually collapse
under its own gravitational force and if $\delta\rho_{m}<0$ it
will expand faster than the average Hubble expansion rate, thus
generating a void. In other words, $\delta\rho_{m}$ is positive in
overdense region and it is negative in underdense regions. In
fact, when the universe is in the matter dominated era, denser
regions expand slower than the entire universe. Therefore if their
density is enough large, they eventually collapse and create
gravitational constraints systems like clusters \cite{Ryden}.
Similar to Eq. (\ref{continuty1}), the conservation equation for
spherical perturbed region can be written as
\begin{equation}
\dot{\rho}_{m}^{c}+3h\rho_{m}^{c}=0, \label{continuty2}
\end{equation}
where $h=\dot{a}_{p}/a_{p}$ is the local expansion rate of the
spherical perturbed region of radius $a_p$ (subscript $p$ refers
to the perturbed). In order to study the evolution of
perturbations, we define a useful and dimensionless quantity
called density contrast as \cite{Ryden}
\begin{equation}
\delta_{m}=\dfrac{\rho_{m}^{c}}{\rho_{m}}-1=\dfrac{\delta\rho_{m}}{\rho_{m}},
\label{delta}
\end{equation}
where $\rho_{m}^{c}$ is the energy density of spherical perturbed
cloud and $\rho_{m}$ is the background density. Taking the
derivative of Eq.(\ref{delta}) with respect to the cosmic time and
using Eq.(\ref{continuty1}) and Eq.(\ref{continuty2}) we obtain
\begin{eqnarray}
&&\dot{\delta}_{m}=3(1+\delta_{m})(H-h),\label{deltadot} \\
&&\ddot{\delta}_{m}=3(\dot{H}-\dot{h})(1+\delta_{m})+
\frac{\dot{\delta}_{m}^{2}}{1+\delta_{m}}, \label{deltadubbledot}
\end{eqnarray}
where the dot denotes the derivative with respect to time.
Combining Eqs. (\ref{Fried1}), (\ref{Fried2}), (\ref{Va}) and
(\ref{lambda}), we arrive at
\begin{equation}
\dfrac{\ddot{a}}{a}=-\dfrac{1}{6}\left(\rho_{m}+\left(\dfrac{-1+3\beta}{1+\beta}\right)\lambda_{0}a^{-3(1+\beta)}\right).
\label{adubbledot}
\end{equation}
According to SC model, a homogeneous sphere of uniform density
with radius $a_{p}$ can itself be modeled using the same equations
that govern the evolution of the universe, with scale factor $a$
\cite{Peebles}. Therefore, we can write for the spherical
perturbed cloud  with radius $a_p$, an equation similar to
Eq.(\ref{adubbledot}), namely
\begin{equation}
\dfrac{\ddot{a}_{p}}{a_{p}}=-\dfrac{1}{6}\left(\rho_{m}^{c}+\left(\dfrac{-1+3\beta}{1+\beta}\right)\lambda_{0}^{c}a_{p}^{-3(1+\beta)}\right),
\label{apdubbledot}
\end{equation}
In general, one may expect $\lambda_{0}$ and $\beta$ differ inside
and outside of the spherical region. However, for simplicity here
we propose they are similar, namely $\lambda_{0}^{c}=\lambda_{0}$
and $\beta^{c}=\beta$. Consider two spheres with equal amounts of
material, one of background material with radius $a$, and one of
radius $a_{p}$ with a homogeneous change in over density, so we
have $\rho^{c}_{m} a_{p}^{3}=\rho_m a^{3}$ \cite{Ryden}. Combining
with Eq.(\ref{delta}), we get
\begin{equation}
a_{p}=a\left( 1+\delta_m\right)^{-1/3}. \label{ap-delta}
\end{equation}
Therefore, Eq.(\ref{apdubbledot}) can be rewritten as
\begin{eqnarray}
&&\dfrac{\ddot{a}_{p}}{a_{p}}=-\dfrac{1}{6}\Bigg{\{}\rho_{m}^{c}+\left(\dfrac{-1+3\beta}{1+\beta}\right)\lambda_{0}
a^{-3(1+\beta)}(1+\delta_{m})^{1+\beta}\Bigg{\}}\nonumber\\
&&=-\dfrac{1}{6}\Bigg{\{}\rho_{m}^{c}+\left(\dfrac{-1+3\beta}{1+\beta}\right)\lambda_{0}
a^{-3(1+\beta)}[1+(1+\beta)\delta_{m}]\Bigg{\}},\nonumber\\
 \label{apduubeldot-delta}
\end{eqnarray}
where we have expanded the last term and only kept the linear term
of $\delta_m$. This is due to the fact that we work in the linear
regime where $\delta_m<1$. Combining Eqs. (\ref{adubbledot}) and
(\ref{apduubeldot-delta}), yield
\begin{equation}
\dot{H}-\dot{h}=\dfrac{1}{6}\delta_{m}\left[\rho_{m}+\left(-1+3\beta\right)\lambda_{0}
a^{-3(1+\beta)}\right]-H^{2}+h^{2}. \label{Hdot-hdot}
\end{equation}
Substituting Eq. (\ref{Hdot-hdot}) into Eq. (\ref{deltadubbledot})
and using Eq. (\ref{deltadot}), we can find the second order
differential equation for the density contrast $\delta_m$ as
\begin{eqnarray}
&&\ddot{\delta}_{m}-\dfrac{1}{2}\left[\rho_{m}+(-1+3\beta)\lambda_{0}a^{-3(1+\beta)}\right]
\delta_{m} \left(1+\delta_{m}
\right)\nonumber\\
&&-\dfrac{4}{3}\dfrac{\dot{\delta}_{m}^{2}}{1+\delta_{m}}+2H\dot{\delta}_{m}=0,
\label{deltaduubeldot2}
\end{eqnarray}
where the matter energy density is given by $\rho_{m}=
\rho_{m,0}a^{-3}$.
In order to study the evolution of the density contrast $\delta_m$
in terms of the redshift parameter, $1+z=1/a$, we first replace
the time derivatives with the derivatives with respect to the
scale factor $a$. It is a matter of calculations to show that
\begin{equation}
\dot{\delta}_{m}=\delta_{m}^{\prime}aH, \quad \ddot{\delta}_{m}
=\delta_{m}^{\prime\prime}a^{2}H^{2}-\dfrac{1}{2}a\left(H^{2}-V\right)\delta^{\prime}_{m},
\label{prim}
\end{equation}
where the prime stands for the derivative respect to $a$.
Therefore Eq.(\ref{deltaduubeldot2}) with using Eq.(\ref{Fried1}) can be written as
\begin{eqnarray}\label{deltafora}
&&\delta^{\prime\prime}_{m}+\dfrac{3}{2a}\delta^{\prime}_{m}+\dfrac{1}{2aH^2}\left( -3H^2+\rho_{m}-\lambda_{0}a^{-3(1+\beta)}\right)\delta^{\prime}_{m}\nonumber\\
&&-\dfrac{1}{2a^2H^2}\delta_{m}(1+\delta_{m})\left( \rho_{m}+(-1+3\beta)\lambda_{0}a^{-3(1+\beta)}\right)\nonumber\\
&&-\frac{4}{3}\dfrac{{\delta^{\prime}_m}^2}{(1+\delta_{m})}=0.
\end{eqnarray}
Since we are working in the linear regime, thus we neglect
$O(\delta_{m}^2)$ and $ O({\delta^{\prime}_m}^2)$. Combining
Eqs.(\ref{Hubble}) and (\ref{deltafora}), we arrive at
\begin{eqnarray}
&&\delta^{\prime\prime}_{m}+\dfrac{3}{2a}\delta^{\prime}_{m}-\dfrac{3}{2a}\beta
\left[ \dfrac{\lambda_{0}a^{-3\beta}}{(\beta+1)\rho_{m,0}-\lambda_{0}a^{-3\beta}}\right]\delta^{\prime}_{m}\nonumber\\
&&-\dfrac{3}{2a^{2}}(\beta+1)\left[
\dfrac{\rho_{m,0}+(-1+3\beta)\lambda_{0}a^{-3\beta}}{(\beta+1)\rho_{m,0}-\lambda_{0}a^{-3\beta}}\right]\delta_{m}=0.\nonumber\\
\label{betadelta}
\end{eqnarray}
It should be noted that either in the absence of the mimetic field
($\lambda_{0}=0$), or in the absence of the mimetic potential
($\beta=0$), Eq. (\ref{betadelta}) reduces to
\begin{eqnarray}\label{deltagr}
&&\delta^{\prime\prime}_{m}+\dfrac{3}{2a}\delta^{\prime}_{m}-\frac{3\delta_{m}}{2a^{2}}=0,
\end{eqnarray}
which is the result obtained in standard cosmology \cite{Abramo}.
This implies that the presence of the mimetic potential plays an
important role in studying the evolution of perturbation in
mimetic cosmology. In other words, in the absence of the mimetic
potential, the perturbed equation for the density contrast,
$\delta_m$, in the linear regime, coincides with the ones in
standard cosmology.
Also in order to express  $\delta_{m}$ as a function of $z$, we
can change the variable from scale factor to the redshift
parameter $z$. In this way, Eq. (\ref{betadelta}) transforms to
\begin{eqnarray}
&&(1+z)^{2}\dfrac{d^{2}\delta_{m}}{dz^{2}}+\dfrac{1}{2}(1+z)\dfrac{d\delta_{m}}{dz}\nonumber\\
&&+\dfrac{3}{2}(1+z)^{(3\beta+1)}\Bigg{\{}\dfrac{\lambda_{0}\beta}{(1+\beta)\rho_{m,0}-\lambda_{0}(1+z)^{3\beta}}\Bigg{\}}\dfrac{d\delta_{m}}{dz}\nonumber\\
&&-\dfrac{3}{2}(1+\beta)\Bigg{\{}
\dfrac{\rho_{m,0}+(-1+3\beta)\lambda_{0}(1+z)^{3\beta}}{(1+\beta)\rho_{m,0}
-\lambda_{0}(1+z)^{3\beta}}\Bigg{\}}\delta_{m}=0.\nonumber\\
\label{generaldelta}
\end{eqnarray}
In this work, we assume $\Omega^{0}_{m}=0.05$,
$\Omega^{0}_{\lambda}=0.26$ and $\Omega^{0}_{V}=0.69$. And thus
from Eqs. (\ref{Omega1}) and (\ref{Omega2}) we have
$\lambda_{0}/\rho_{m,0}=\Omega^{0}_{\lambda}/\Omega^{0}_{m}=+5.2$,
($\lambda_0>0$). Therefore, Eq. (\ref{generaldelta}) admits the
following analytical solution
\begin{eqnarray}\label{solution}
&&\delta_{m}(z)= \Bigg{\{}C_{1}(1+z)^{\alpha_{+}}
{}_2F_1\left(\eta_{+},\eta_{-},s_{+},\frac{1923(1+\beta)}{10000(1+z)^{3\beta}}\right)\nonumber\\
&&+C_{2} (1+z)^{\alpha_{-}}\times
{}_2F_1\left(\eta_{+},\eta_{-},s_{-},\frac{1923(1+\beta)}{10000(1+z)^{3\beta}}\right)\Bigg{\}}M,\nonumber\\
\end{eqnarray}
where $C_{1}$ and $C_{2}$ are integration constants and
\begin{eqnarray} \label{parameter}
&& M=\left[-19230000+(3697929\beta+ 3697929)(1+z)^{-3\beta} \right]^{1/4}\nonumber\\
&&\times\left[-10000+(1923+1923\beta)(1+z)^{-3\beta} \right]^{5/4},\nonumber\\
&&\alpha_{\pm}=\dfrac{1}{4}\pm\dfrac{1}{28}\sqrt{7}\sqrt{-175+294\beta+441\beta^{2}}+\dfrac{3}{4}\beta,\nonumber\\
&&\eta_{\pm}=-\dfrac{1}{84\beta}\Big{\{}-105\beta\mp35+\sqrt{7}\sqrt{-175+294\beta+441\beta^{2}}\Big{\}},\nonumber\\
&&s_{\pm}=\mp\dfrac{1}{42\beta}\Big{\{}\mp42\beta+\sqrt{7}\sqrt{-175+294\beta+441\beta^{2}}\Big{\}}.
\end{eqnarray}
To obtain the integration constants, we note that for large values
of the redshift parameter, the parameter $\beta$ goes to zero
where the GR limit is recovered. We therefore consider the
adiabatic initial conditions for matter perturbations as
\cite{Abramo}
\begin{equation}
\dfrac{d\delta_{m}(z)}{dz}(\mid_{z=z{i}})=-\dfrac{\delta_{m}(z_{i})}{1+z_{i}},
\label{initial condition}
\end{equation}
where $\delta_{m}(z_{i})$ is the initial value for the density
contrast at $z=z_{i}$. In Fig. 8, we have plotted the matter
density contrast as a function of redshift for different values of
$\beta$ parameter and for redshifts $10<z<100$. We observe
that in the framework of mimetic gravity, the growth of matter
perturbations is larger than $\Lambda CDM$ model. Indeed, the
density contrast of matter starts growing from its initial value
and as the universe expands, the matter density contrast grows up
faster and deviates from the $\Lambda CDM$ profile. We find out
that $\beta$ parameter affects the growth of matter perturbations,
in particular in the lower redshifts. In fact, the growth of
perturbations increases with increasing $\beta$, which reveals the
influences of the mimetic field. The physical reason behind this
behavior comes from the fact that the mimetic field encodes an
extra longitudinal degree of freedom to the gravitational field,
which can supports the growth of matter perturbations.
\begin{figure}[t]
\epsfxsize=8.5cm \centerline{\epsffile{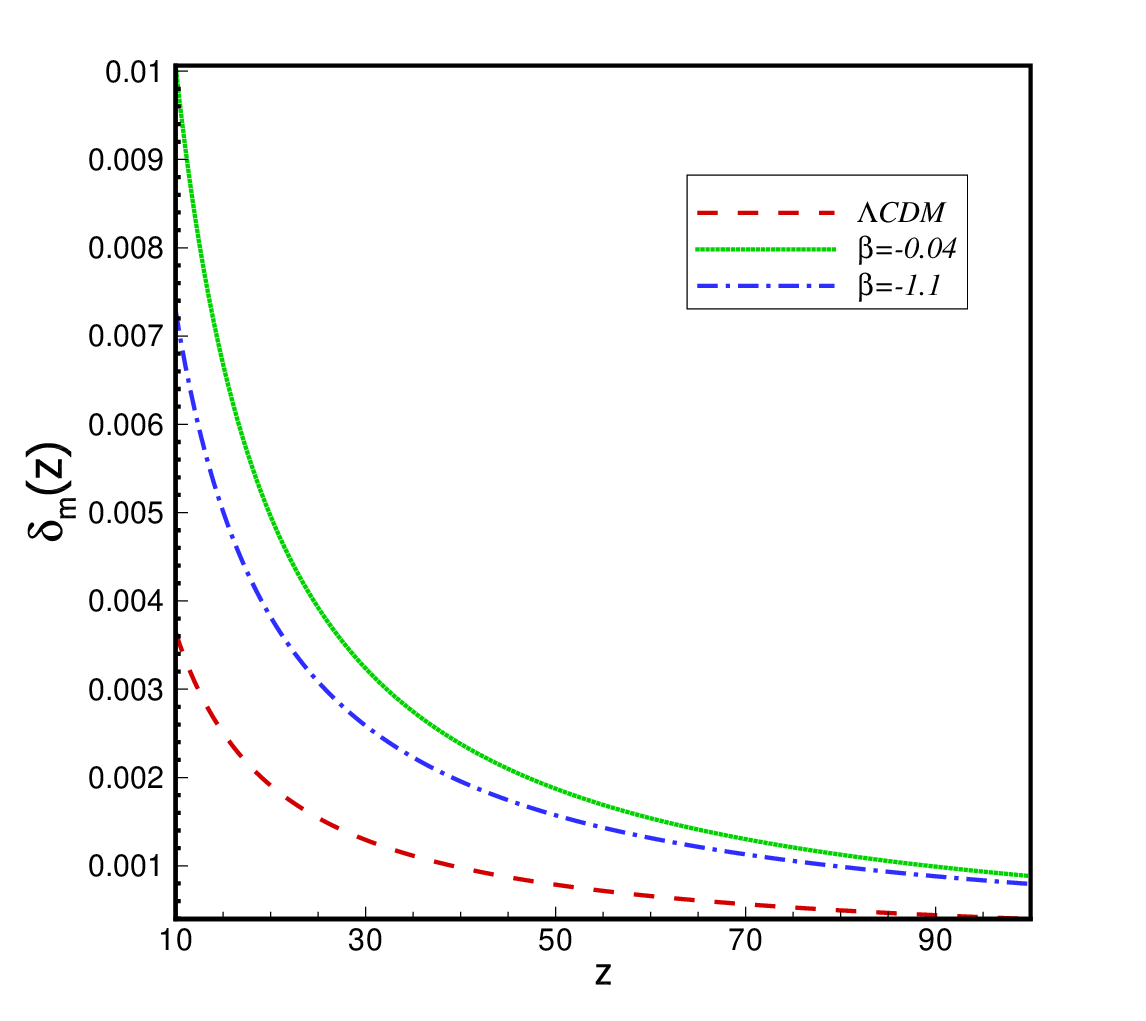}} \caption{The
evolution of the matter density contrast for different values of
$\beta$ during the evolution of the Universe. We have chosen
$\delta_{m}(z_{i}) =0.0001$ and $z_{i}=400$.} \label{fig8}
\end{figure}
We remind that the universe goes through several phase
transitions during its formative stages. Cosmic reionization is the
last of them, where ultraviolet and X-ray radiation escape from
the first generations of galaxies heating and ionizing their surroundings
and subsequently the entire intergalactic medium \cite{Wise}. This occurred
between 150 million and one billion years after the Big Bang
($6<z<20$) \cite{Wise}. The results suggest that the universe
was approaching the end of reionization around z=6. It suggests that
the universe must still have been almost entirely neutral at $z>10$.
This is the physical reason that we take the perturbations for $10<z<100$ in Fig. 8. \cite{Wise}.\\
We can investigate the growth rate of matter perturbations which
is given by the growth function as \cite{Peebles}
\begin{equation}\label{ growth rate}
f(a)=\dfrac{dlnD}{dlna},\      \     \     \    \
D(a)=\dfrac{\delta_{m}(a)}{\delta_{m}(a=1)}.   
\end{equation}
Let us note that in the absence of the mimetic field
($\lambda_{0}= 0$), or in the absence of the mimetic potential
($\beta=0$), the growth function is a constant of unity. In Fig. 9
we have plotted the growth function in terms of the redshift
parameter. We observe the amplitude of perturbations in
high redshifts corresponds to the unity, but it starts to decrease
at low redshifts similar to $\Lambda CDM$ model. Therefore, the
role of potential (DE), like $\Lambda$, is to reduce the growth
function from unity. Besides, the current value of $f(z)$
crucially depends on the $\beta$ parameter and decreases with
decreasing $\beta$.
\begin{figure}[t]
\epsfxsize=6.8cm \centerline{\epsffile{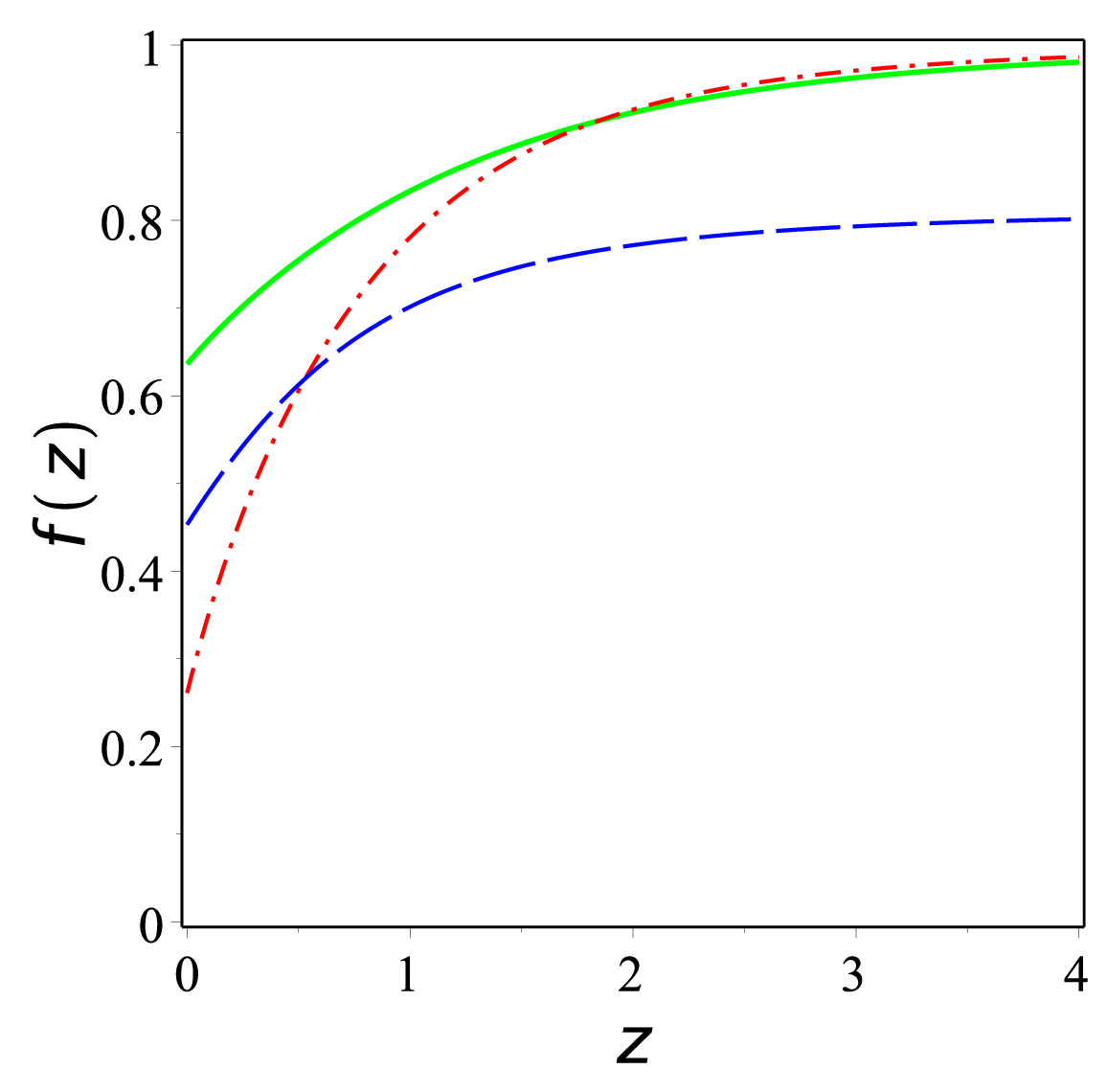}} \caption{The
evolution of the growth function for different values of $\beta$ parameter,
where solid-line for $\beta=-1.1$, dashdotted-line for $\beta=-1.2$, long dashed-line for $\Lambda CDM$.}
\label{fig9}
\end{figure}\\
We can also measure the growth rate matter perturbations from the
redshift-space distortion (RSD) of clustering pattern of galaxies.
This distortion is caused by the peculiar velocity of inward
collapse motion of the large-scale structure, which is directly
related to the growth rate of the matter density contrast
$\delta_{m}$ \cite{Kaiser}. Recent galaxy redshift surveys have
provided bounds on the growth rate $f(z)$ or $f(z)\sigma_{8}(z)$
in terms of the redshift where $f(z)$ is from Eq. (\ref{ growth
rate} ) and $\sigma_{8}$ is the rms amplitude of $\delta_{m}$ at
the comoving scale $8h^{-1}Mpc$ \cite{Tsujikawa, Nesseris} and can
be written as \cite{Nesseris2}
\begin{equation}
\sigma_{8}(z)=\dfrac{\delta(z)}{\delta(z=0)}\sigma_{8}(z=0),
\label{sigma8z}
\end{equation}
where we have assumed $\sigma_{8}(z=0)=0.983$ \cite{Nesseris}, and
we have shown the redshift evolution of $f(z)\sigma_{8}(z)$ for
different values of $\beta$ parameter in Fig. 10. We see that the
growth rate of models with potential (nonzero $\beta$) is smaller
than model without potential. Although the behaviour of models
with potential are similar for large redshifts but in small
redshifts ($z<1$) the model with larger $\beta$ parameter predicts
larger value of the cosmological growth rate. From Fig. 10
we observe that in small redshifts, the growth rate function with
potential is smaller than $\Lambda CDM$ model. Also one can see
that the evolution of the function $f \sigma_{8}$ for
$\beta=-1.2$, is very similar to the $\Lambda CDM$ theory. The
difference can be traced back to the linear regime of the
solutions. The present value of $f \sigma_{8}$ for $\beta=-1.14$
is greater than of the $\Lambda CDM$ value. Also we see for larger
value of $\beta$ parameter $f \sigma_{8}$ reaches the maximum
value at smaller redshifts. In other words, with increasing the
role of potential (DE) the large structures has been formed later
in this mimetic gravity.
\begin{figure}[t]
\epsfxsize=7.5cm \centerline{\epsffile{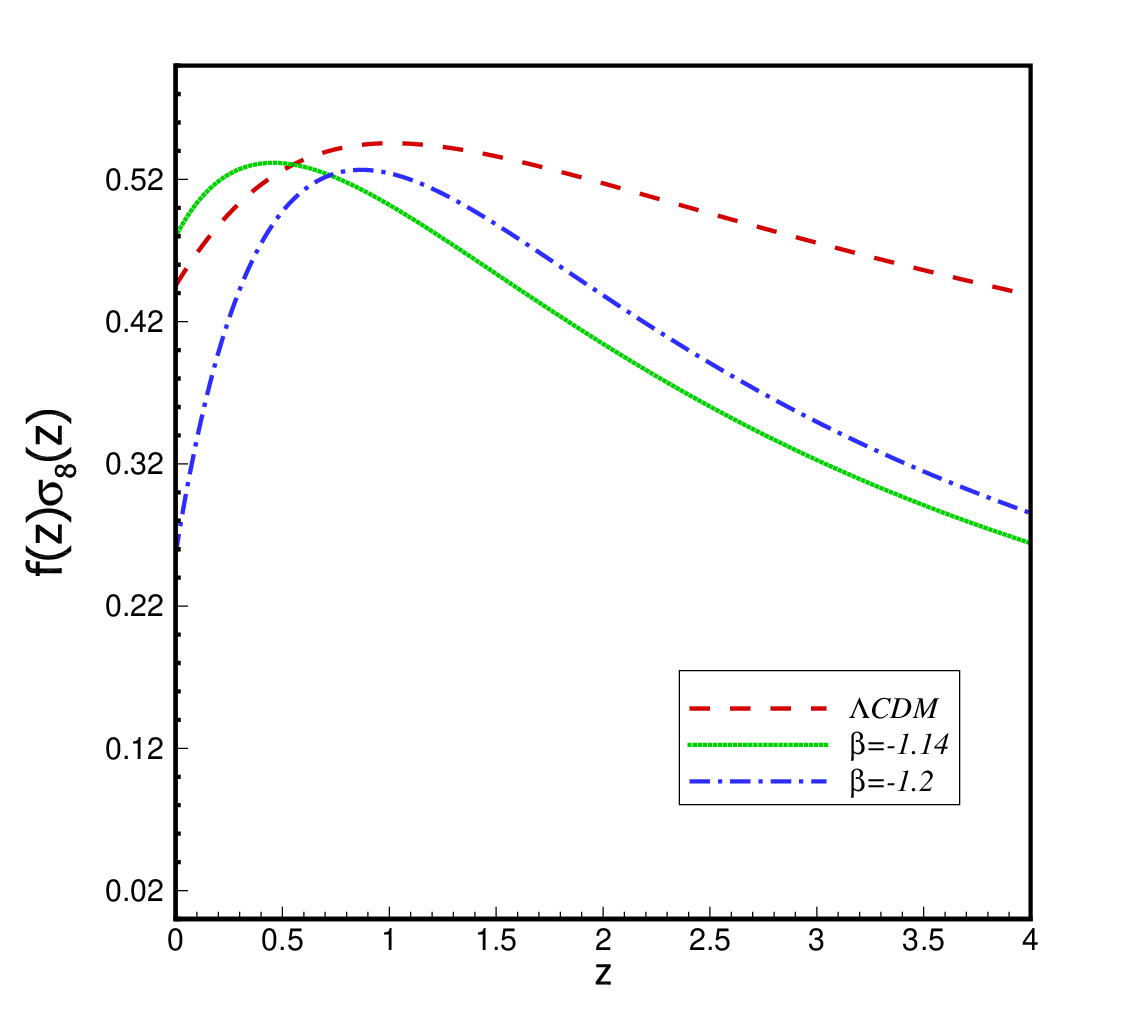}} \caption{The behavior of $f(z)\sigma_{8}(z)$ for different values of $\beta$ parameter.} \label{fig10}
\end{figure}
\section{Mass function and cluster number counts\label{MN}}
Besides the evolution of matter density contrast, in the
present work the number count of the collapsed objects in the
mimetic scenario is also emphasized. The gravitational collapse of
the matter overdensity is the basic process of large scale
structure formation in the universe. The objects, formed by the
collapse, are called the DM halos. Due to gravitational attraction
the baryonic matter follows the distribution DM. Hence the galaxy
clusters are embedded in the DM halos. The observed distribution
of galaxy clusters provides the information about the distribution
of DM halos in the universe. In this section, using the SC model,
we study the DM halo mass function and the number distribution in
the context of mimetic cosmology.

Now we want to use the Press-Schechter approach in order to
estimate the number counts of DM halos for different bins of
redshifts and halo masses \cite{William}. The Press-Schechter
formalism assumes the fraction of mass in the universe contained
in gravitationally bound systems with masses greater than $M$ is
given by the fraction of space where the linearly evolved density
contrast exceeds a threshold $ \delta_{c}$, and that the density
contrast is normally distributed with zero mean and variance
$\sigma^{2}(M)$ -the root-mean-square value of the density
contrast $\delta$ at scales containing mass $M$. Therefore, it is
assumed that for a massive sphere to undergo gravitational
collapse at a redshift $z$ its linear overdensity should exceed a
threshold $\delta_{c}(z)$. Notice that only linear quantities are
used in this formalism. The key quantity in the SC is the
critical overdensity $\delta_{c}(z_{c})$ at a given collapse redshift $z_{c}$.
It is defined as the value of the linear density
contrast at the redshift where the nonlinear density contrast
diverges. In other words it is defined as the final value (i.e, at redshift $z_{c}$) of the
linear evolution of a given spherical top-hat initial perturbation that
actually collapses at $z_{c}$ according to the full nonlinear equations.

These assumption lead to the well-known analytical formula for the comoving
number density of collapsed halos of mass in the range $M$ and $M+dM$
at a given redshift $z$ \cite{Liberato}:
\begin{equation}
\dfrac{dn}{dM}=-\sqrt{\dfrac{2}{\pi}}\dfrac{\rho_{m,0}}{M}
\dfrac{\delta_{c}(z)}{\sigma(M,z)}\dfrac{d\ln\sigma(M,z)}{dM}\exp\left[ -\dfrac{\delta_{c}^{2}(z)}{2\sigma^{2}(M,z)}\right],
\label{massfunction}
\end{equation}
where $\rho_{m,0}$ is the peresent matter mean density of the
universe and $\delta_{c}(z)$ is the linearly extrapolated density
threshold above which structures collapse, i.e.
$\delta_{c}(z)=\delta(z_c)$. In an Einstein-de Sitter (EdS)
universe, an overdensity region collapses with a linear contrast
$\delta_{c}=1.686$ \cite{Padmanabhan}. In order to obtain
$\delta_c$ in terms of the redshift parameter $z$, we consider
linear equation (\ref{betadelta}) for the spherical region where
$\delta_m=\delta_c$. This equation with using the
differential-radius method \cite{Herrera} has the solution in the
form $\delta_{c}(a)\propto a_{c}^{(A+3)/4}$, in far redshifts. In
other words with $1+z\propto t^{-2/3}$, we can see the time
dependence $\delta_{c}\propto t^{m}$, where
$m=(\sqrt{1+24\Omega_{m}}-1)/6$ \cite{Nunes}. Therefore, its
solution can be written as
\begin{equation}
\delta_{c}(z)=\dfrac{\delta_{i}}{2}\left(1+\dfrac{1}{A}\right)
\left(\dfrac{1+z{i}}{1+z}\right)^{(A-1)/4}, \label{deltacrit}
\end{equation}
where $A=\sqrt{25-72\beta}$ depends on the model parameter
$\beta$. Let us note that for an initial perturbation
$\delta_{i}=10^{-7}$ and $1+z_{i}=10^{8}$, and the overdensity
region collapses at redshift $z_{c}=2.558$,  Eq. (\ref{deltacrit})
gets $\delta_{c}=1.686$, which corresponds to the Eds value. The
quantity
\begin{equation}
\sigma(M,z)=D(z)\sigma_{M},
\label{sigmam,z}
\end{equation}
is the linear theory "rms" density fluctuation in spheres of comoving radius
$R$ containing the mass $M$, where $D(z)\equiv\delta_{m}(z)/\delta_{m}(z=0)$
is the linear growth function obtained from Eq. (\ref{ growth rate} ). The smoothing
scale $R$ is often specified by the mass within the volume defined by the window
function at the present time, see e.g. \cite{Peebles}. In our analysis the variance
of the smoothed overdensity containing a mass $M$ is given by \cite{Viana}:
\begin{equation}
\sigma_{M}=\sigma_{8}\left( \dfrac{M}{M_{8}}\right)^{-\gamma(M)/3},
\label{sigmam}
\end{equation}
where $M_{8}=6\times 10^{14}\Omega_{M}^{(0)}h^{-1}M_{\bigodot}$ is
the mass inside a sphere of radius $R_{8}=8h^{-1}Mpc$, and $\sigma_{8}$
is the variance of the over-density field smoothed on a scale of size $R_{8}$ \cite{Viana}.
The index $\gamma(M)$ is a function of the mass scale and the shape parameter,
$\Gamma$, of the matter power spectrum \cite{Viana}
\begin{equation}
\gamma=(0.3\Gamma+0.2)\left[ 2.92+\dfrac{1}{3} \log\left(  \dfrac{M}{M_{8}}\right) \right].
\label{gamma}
\end{equation}
In our study we use $\Gamma=0.167$ \cite{Spergel}. \\
Denoting $\tilde{\gamma}(M)=d ln\sigma(M,z)/dM$,
\begin{equation}
\tilde{\gamma}(M)=(0.3\Gamma+0.2)\left[ 2.92+\dfrac{2}{3} \log\left(  \dfrac{M}{M_{8}}\right) \right],
\label{gammatilde}
\end{equation}
we can rewite Eq.(\ref{massfunction}) as
\begin{equation}
\dfrac{dn}{dM}=-\sqrt{\dfrac{2}{\pi}}\dfrac{\rho_{m,0}}{M}\dfrac{\delta_{c}(z)}{\sigma(M,z)} \tilde{\gamma}(M) exp\left[ -\dfrac{\delta_{c}^{2}(z)}{2\sigma^{2}(M,z)}\right],
\label{massfunction2}
\end{equation}
For a fixed $\sigma_{8}$ (power spectrum normalization) the
predicted number density of DM halos given by the above formula is
uniquely affected by the DE models, with different $\beta$
parameter, through the ratio $\delta_{c}/D(z)$.
\begin{figure}[t]
\epsfxsize=7.5cm \centerline{\epsffile{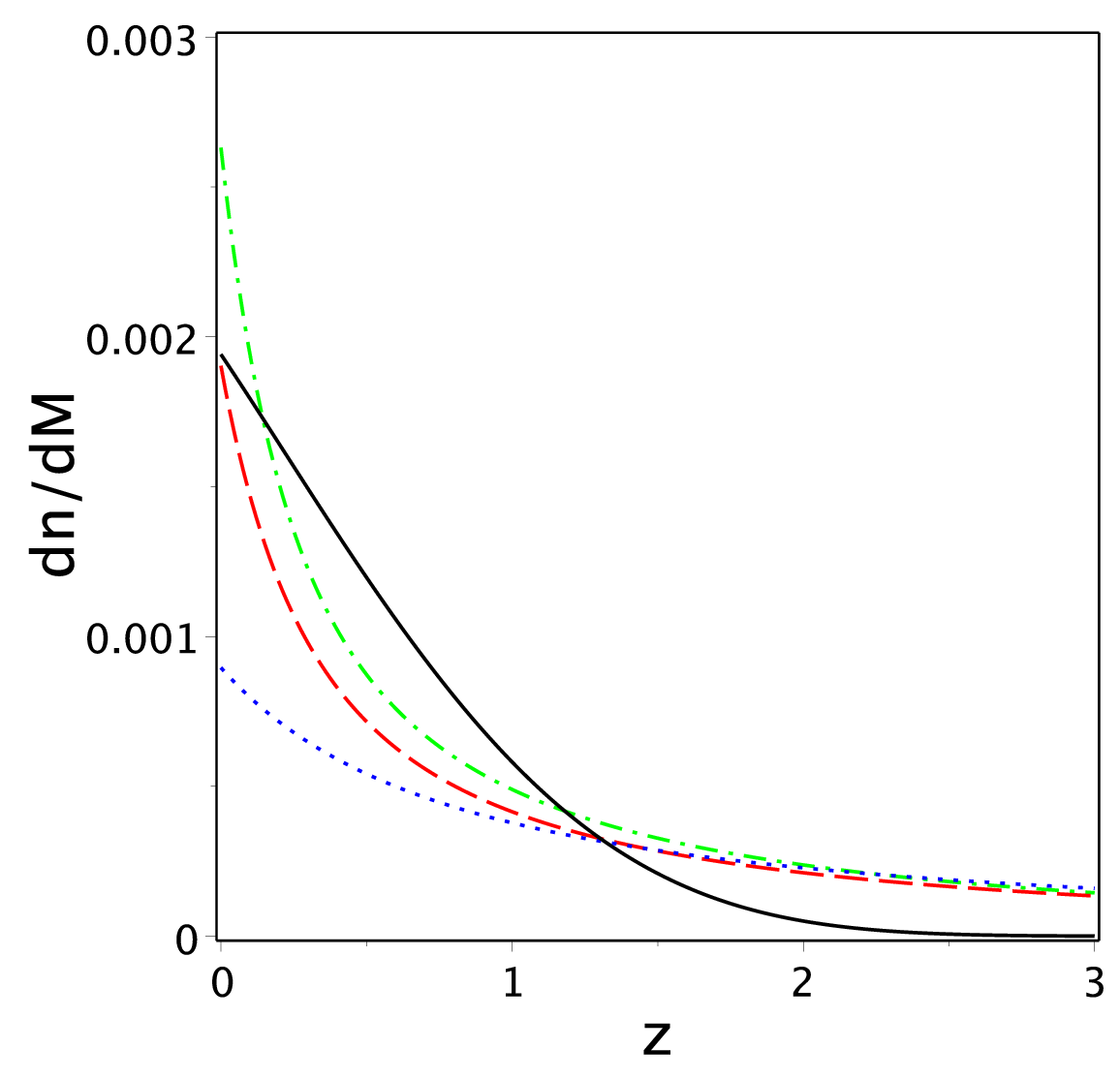}} \caption{The evolution of mass function for objects with mass $M=10^{13}(h^{-1}M_{\bigodot})$ with different $\beta$ parameter, where dashdotted-line for $\beta=-1.6$, dashed-line for $\beta=-1.4$, dotted-line for $\beta=-1.01$ and solid-line for$\Lambda CDM$. }
\label{fig11}
\end{figure}
In Fig. 11, we have shown the redshift evolution of the mass
function ($dn/dM (1/Mpc^3)$) of objects with mass $10^{13}
h^{-1}M_{\bigodot}$ for different $\beta$ parameter. From this
figure, we can investigate the modifications caused by a DE
component on the number of DM halos. We observe that the
mass function of models with potential (nonzero $\beta$) starts to
grow around $z\sim2$. Besides, the mass function for
$\beta=-1.01,-1.4$ are smaller than $\Lambda CDM$ model, while for
$\beta=-1.4$ its present value correspond to $\Lambda CDM$ model.
Also the smaller value of $\beta$ parameter give larger halo
abundances. In other words from Fig. 11, we find out that by
reducing the role of DE, then the comoving number density becomes
more. We can see that with decreasing the role of DE (for smaller
$\beta$ parameter), the mass function starts to grow in smaller
redshifts i.e., halo abundance is formed later. Also we find this
same qualitative behaviour within the mass range, $10^{13}-10^{16}
h^{-1}M_{\bigodot}$, for dark energy models with different $\beta$
parameter.

The effect of dark energy on the number of DM halos is studied by
computing two quantities. The first is all sky number of halos per
unit of redshift, in a given mass bin \cite{Nunes}
\begin{equation}
\mathcal{N}^{bin}\equiv\dfrac{dN}{dz}=\int_{4\pi} d\Omega \int^{M_{sup}}_{M_{inf}}\dfrac{dn}{dM}\dfrac{dV}{dzd\Omega}dM,
\label{numbercounts}
\end{equation}
where the comoving volume element is given by
\begin{equation}
\dfrac{dV(z)}{dz d\Omega}=\dfrac{r^{2}(z)}{H(z)},
\label{volume}
\end{equation}
where $r(z)=\int^{z}_{0} H^{-1}(x)dx$ is the comoving distance. We
have plotted the redshift evolution of the comoving volume element
($dV/dz d\Omega(Mpc^3)$) with different $\beta$ parameter in Fig.
12. As we see the comoving volume element is larger for smaller
values of $\beta$ parameter. Note that the comoving volume element
does not depend on the growth factor of the perturbation $D(z)$,
but only on the cosmological background.
\begin{figure}[t]
\epsfxsize=7.5cm \centerline{\epsffile{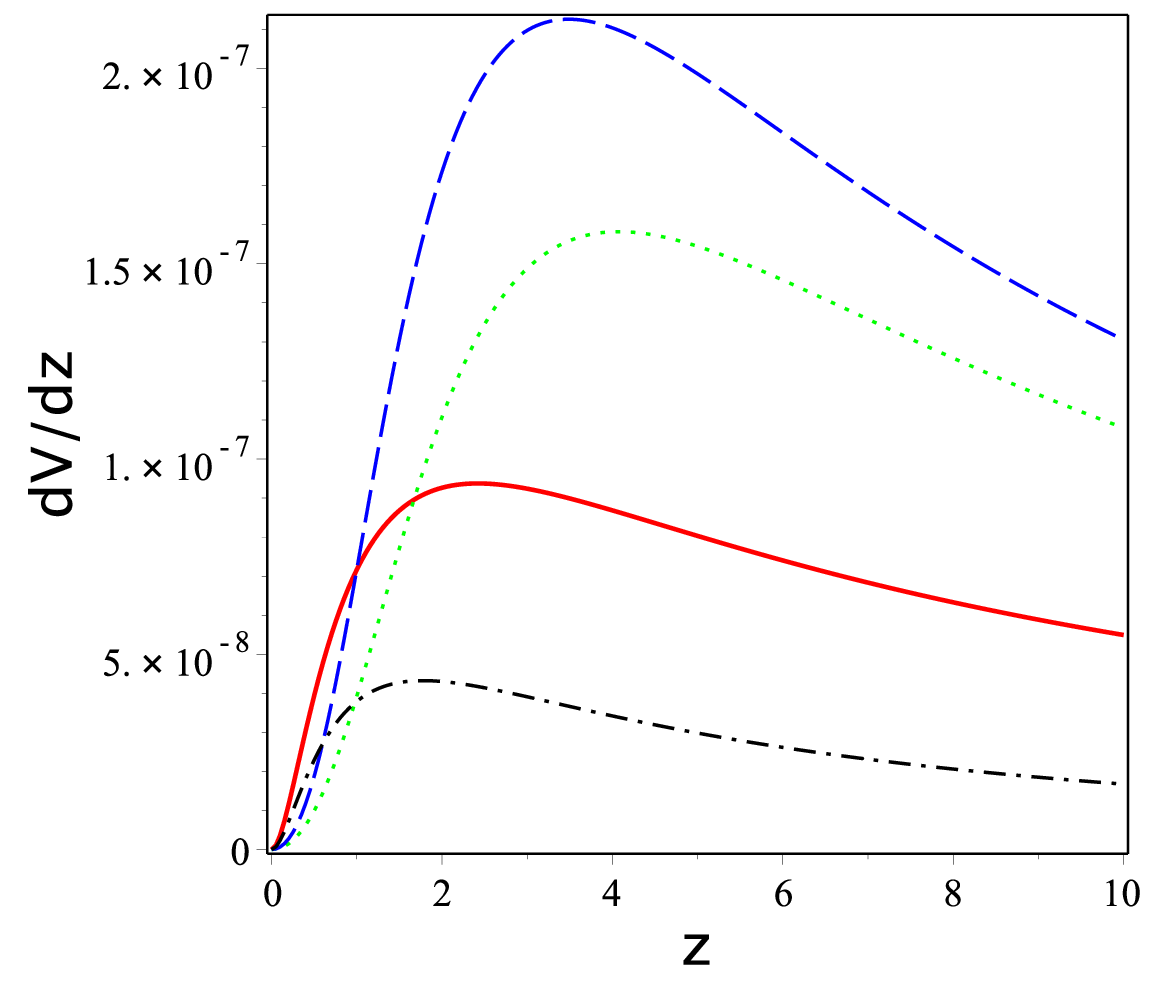}} \caption{The
behavior of the comoving volume element with different $\beta$
parameter, where dashed-line for $\beta=-1.4$, dotted-line for $\beta=-1.2$, solid-line for
$\beta=-0.09$, dashdotted-line shows it for $\beta=0$.} \label{fig12}
\end{figure}
The number counts in mass bins, $\mathcal{N}^{bin}=dN/dz$,
obtained from Eq. (\ref{numbercounts}), is shown in Fig. 13. In
this figure we show it for $\beta=0$ model and for different mass
bins $[M_{inf},M_{sup}]$ as $[10^{12},10^{13}]h^{-1}M_{\bigodot}$,
$[10^{13},10^{14}]h^{-1}M_{\bigodot}$,
$[10^{14},10^{15}]h^{-1}M_{\bigodot}$. As we see the more massive
structures are less abundant and form at later times, as it should
be in the hierarchical model of structure formation.
\begin{figure}[t]
\epsfxsize=7.5cm \centerline{\epsffile{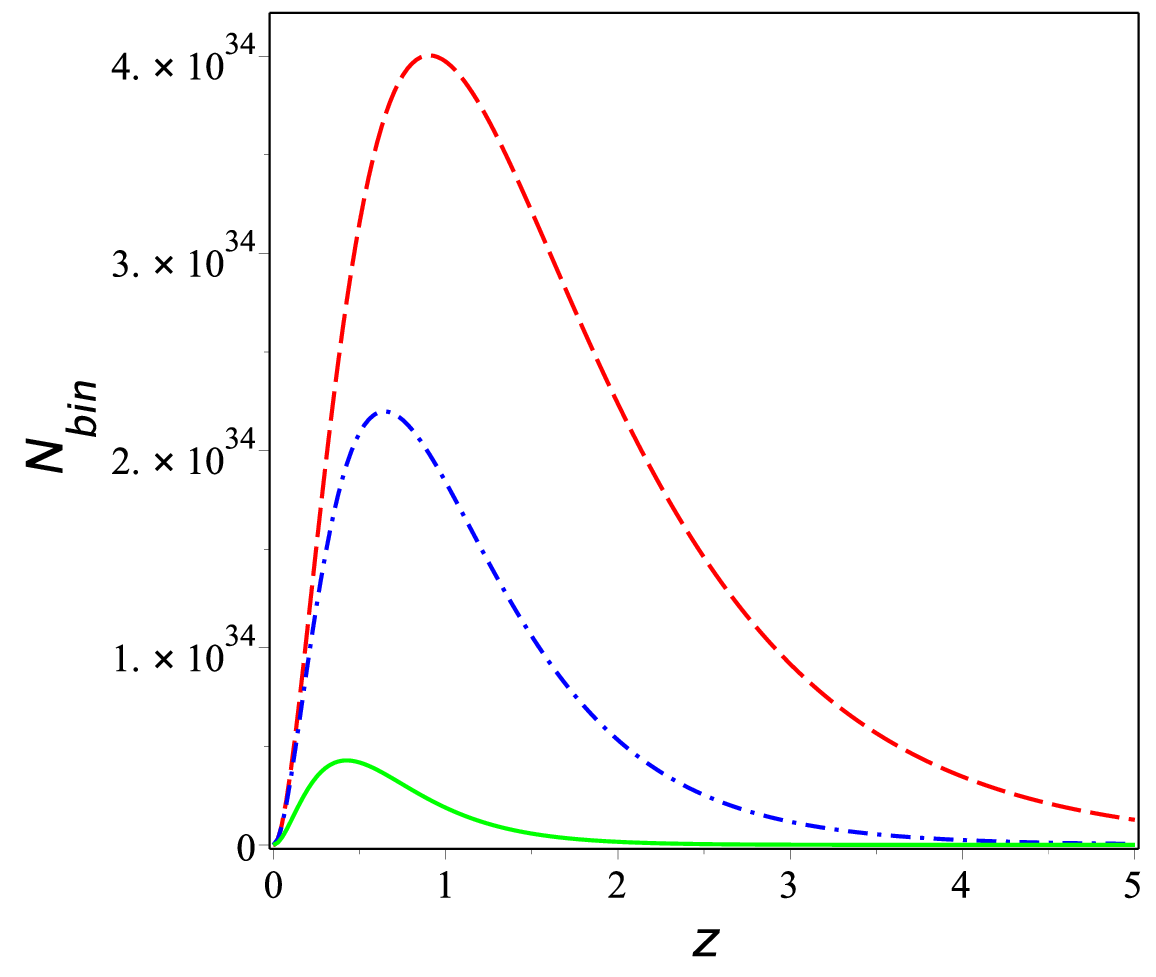}} \caption{The
Evolution of number counts with redshift from the $\beta=0$ model
for objects with mass within the range, dashed-line for $10^{12}<M/(h^{-1}M\bigodot)<10^{13}$,
 dashdotted-line for $10^{13}<M/(h^{-1}M\bigodot)<10^{14}$, solid-line for $10^{14}<M/(h^{-1}M\bigodot)<10^{15}$.} \label{fig13}
\end{figure}
Also the second quantity is the all sky integrated number counts
above a given mass threshold, $M_{inf}$ and up, to redshift
\cite{Nunes}
\begin{equation}
\mathcal{N}(z,M>M_{inf})=\int_{4\pi}d\Omega \int^{\infty}_{M_{inf}} \int^{z}_{0} \dfrac{dn}{dM}\dfrac{dV}{dz^{\prime} d\Omega}dMdz^{\prime}
\label{integratednumbercounts}
\end{equation}
\section{Conclusion and discussion\label{CD}}
The discovery that the expansion of the Universe is accelerating
has led to large observational programs being carried out to
understand its origin. New facilities are being designed and built
aiming to measure the expansion history and the growth of
structure in the Universe with increasing precision out to greater
redshifts. Since the dark sectors changes the expansion history of
the Universe and the evolution of matter density perturbations,
these new facilities will not only test the current period of
accelerated expansion but also explain the nature of the dark
sectors.

In this paper, we have investigated the evolution of the matter
perturbations in the context of mimetic gravity for different
values of the model parameters. We have employed the SC formalism
in order to examine the perturbations and worked in the linear
regime. We proposed the mimetic potential (which plays the role of
DE) is proportional to the Mimetic DM density, i.e., $V\propto
\lambda$. With this ansatz, we found out $\lambda\sim
a^{-3(1+\beta)}$, implying an interaction between DE and DM in
this model. The density contrast has similar behavior for
different values of $\beta$ parameter; that is, it starts from its
initial value and then it grows up at low redshifts. Besides, the
density contrast grows faster for larger values of $\beta$
parameter. Furthermore, as the value of the $\beta$ parameter
decreases, then the growth of matter perturbations is decreasing,
too. In other words we see that the growth rate of models with
potential (nonzero $\beta$) is smaller than model without
potenial. Also we studied the modifications caused by a dark
energy component on the number of DM halos as the smaller value of
$\beta$ parameter give larger halo abundances. Also we see that
for smaller $\beta$ parameter and therefore with decreasing the
role of DE, the mass function start to grow in smaller redshifts
i.e. halo abundance is formed later. Also we see the comoving
volume element is larger for smaller values of $\beta$ parameter.
In addition we found the number counts in mass bins for the more
massive structures are less abundant and form at later times, as
it should be in the hierarchical model of structure formation. We
also studied the evolution of the density parameters. We observed
that the evolution of the matter density abundance and the DM
density abundance for different values of $\beta$, decrease by
decreasing the redshift. We found out that the density abundance
of matter, consist of baryonic and DM, drops slower for larger
values of $\beta$. From the evolution of deceleration parameter,
we see that the Universe experiences a phase transition from
decelerating phase to accelerating phase around redshift
$z_{tr}\approx0.6$ which is compatible with observational data.

It is interesting to constrain the model presented here by using
observational data, such as observations of type Ia supernova and
Baryon Acoustic Oscillations (BAO). We leave this issue for future
investigations.

\acknowledgments{We are grateful to the anonymous referees for
very helpful and constructive comments which helped us improve our
paper significantly. We also thank H. Moradpour and M. Asghari for
useful discussion and valuable comments. }

\end{document}